\documentclass[galaxies,review,accept,pdftex,oneauthor]{Definitions/mdpi} 
\firstpage{1} 
\pubvolume{11}
\issuenum{2}
\articlenumber{40}
\pubyear{2023}
\copyrightyear{2023}
\externaleditor{Academic Editors: Dominic Bowman, Jennifer Van Saders and Jorick \linebreak S. Vink 
 }
\datereceived{24 January 2023} 
\daterevised{24 February 2023} 
\dateaccepted{1 March 2023} 
\datepublished{5 March 2023} 
\hreflink{https://doi.org/10.3390/\linebreak galaxies11020040} 

\usepackage{physics}
\usepackage{natbib}
\usepackage{fancyhdr}
\usepackage{amsmath}

\Title{Magnetism in High-Mass Stars}

\TitleCitation{Magnetism in High-Mass Stars}

\Author{{Zsolt Keszthelyi} 
 $^{1,2}$\orcidA{}}

\AuthorNames{Zsolt Keszthelyi}

\AuthorCitation{Keszthelyi, Z.}

\address{%
$^{1}$ \quad Center for Computational Astrophysics, Division of Science, National Astronomical Observatory of Japan, 2-21-1, Osawa, Mitaka, Tokyo 181-8588, Japan; email: zsolt.keszthelyi@nao.ac.jp
\\
$^{2}$ \quad Anton Pannekoek Institute for Astronomy, Faculty of Science, University of Amsterdam, Science Park 904, \linebreak  1098 XH, Amsterdam, The Netherlands
}

\abstract{Magnetism is a ubiquitous property of astrophysical plasmas, yet stellar magnetism still remains far from being completely understood. In this review, we describe recent observational and modelling efforts and progress to expand our knowledge of the magnetic properties of high-mass stars. Several mechanisms (magneto-convection, mass-loss quenching, internal angular momentum transport, and magnetic braking) have significant implications for stellar evolution, populations, and end-products. Consequently, it remains an urgent issue to address and resolve open questions related to magnetism in high-mass stars.}


\begin{document}



\section{Introduction}

Magnetism plays an important role in several astrophysical phenomena. For example, stellar and planetary magnetic fields may be a crucial aspect of planetary habitability, e.g.,~\cite{vidotto2013,varela2022}. Magnetic fields may significantly impact star and planet formation, e.g.,~\cite{commercon2011,Seifried2020a}. 
The magnetic characteristics of neutron stars have been extensively studied via observations, e.g.,~\cite{kaspi2017,enoto2019} and numerical approaches, e.g.,~\cite{reisenegger2009,beloborodov2009}, and the magnetic properties of the interstellar medium, galaxies, and galaxy clusters have also been subject to ongoing investigations, e.g.,~\cite{akahori2018}.
%
However, the magnetic properties of non-degenerate stars are still not fully understood. Cowling and Spitzer concluded that, based on the available magnetic flux during star formation, it should be possible that some remnant, so-called fossil fields, remains in stars~\cite{cowling1945,spitzer1958}. Such fossil fields will be discussed in detail in this review. 
Moreover, the dynamical characteristics of stars also make them plausible sites to generate in situ magnetic fields. A contemporaneously generated (and re-generated) magnetic field taps from a kinetic energy reservoir and converts it to magnetic energy. This is the fundamental principle of a dynamo. Various dynamo processes have also been proposed to operate in stars, and we will address this topic too.
In this review, we will focus on \emph{high-mass stars} with O and B spectral types. They are characterised by a convective core and a radiative envelope on the main sequence. This structural property is fundamental to unveil their magnetic characteristics. 
High-mass stars play a vital role in the chemical evolution of the Universe, e.g.,~\cite{nomoto2013}. Via their powerful stellar winds and ionising radiation, they regulate star and planet formation by stellar feedback, e.g.,~\cite{geen2021, kim2021}.
Stars initially more massive than about 8~M$_\odot$ leave behind neutron stars and black holes by the end of their evolution. Understanding the structure and evolution of the progenitors of compact remnants is required to explain several astrophysical phenomena, which range from observations of fast-radio bursts, e.g.,~\cite{boc2020,chime2020,petroff2022} to detections of gravitational wave events, e.g.,~\cite{abbott2016}.

\newpage
\section{Observations}

In this section, we briefly address some milestones from observational studies as well as important new clues that help us shape theoretical advancements. For a comprehensive background on observational aspects applicable to various kinds of stars, including the Zeeman effect, polarimetry, and modern spectral convolution and imaging techniques, we refer the reader to~\cite{deltoro2003,donati2009,kochukhov2021b}.

\subsection{Historical Perspective}

The first confirmation that magnetism is present in stars took place shortly after Zeeman's discovery of spectral line splitting under external magnetic fields. Hale could measure the line splitting in sunspots and infer their magnetic field strengths~\cite{hale1908}. 
Since then, our knowledge of solar activity has significantly improved, and it was understood that the underlying origin of the observed magnetism is related to convection and rotation, e.g.,~\cite{parker1958,weiss2000}, although the exact role of convection is still debated, e.g.,~\cite{spruit2002}.
In fact, while there still remain open questions about the solar cycle and the solar magnetic field generation and activity, given the supporting observational evidence, it is now generally accepted that all solar-type stars (and, more generally, stars with convective surfaces, including M dwarfs, Red Giants, and Red Supergiants) produce a dynamo mechanism given the ingredients of convection and differential rotation, e.g.,~\cite{mohanty2003,kochukhov2021b,mathias2018}. 
This significant finding further raises the question of whether any convectively unstable region in stars could operate a dynamo cycle and thus produce magnetic activity hidden from the observer.

\subsection{From Ap/Bp to OBA Stars}\label{sec:magobs}

The first magnetic field on a star other than the Sun was detected by Babcock on a chemically peculiar star, 78 Vir~\cite{babcock1947}. In subsequent years, the discovery of chemically-peculiar Ap/Bp stars grew further, e.g.,~\cite{landstreet1977,borra1980}. It remains a very rich field of study, with increasingly more sophisticated techniques to infer surface spots and abundance patterns, e.g.,~\cite{huemmerich2020,giarrusso2022,hajduk2022,kobzar2022,semenko2022}.
There is a demonstrable link between the magnetic characteristics of magnetised pre-main sequence Herbig Ae/Be stars and main sequence Ap/Bp stars~{\cite{alecian2013,villebrun2019,lavail2020,kholtygin2021}}, suggesting that the observed magnetic fields of main sequence stars are already developed in the pre-main sequence phase. This observational evidence is a crucial piece of information in ongoing investigations about the origin of fossil fields.

One indirect hint that other early-type main-sequence stars with radiative envelopes may harbour surface magnetic fields originated from the X-ray properties and evidence of material falling back onto the stellar surface of $\Theta^1$~Ori~C, e.g.,~\cite{conti1972,stahl1996,wade2006}. Analytical wind-shock models were developed in the context of a stellar \textit{magnetosphere} ~\cite{babel1997}. These predictions were also verified by state-of-the-art simulations studying the interaction of stellar winds and the magnetosphere~\cite{ud2002}, as well as further X-ray observations~\cite{gagne2005}. The X-ray characterisation of magnetic massive stars remains invaluable, e.g.,~\cite{cohen2003,naze2014,fletcher2018}.
Importantly, these predictions also led to optical spectropolarimetric observations (using circular polarisation, that is, the Stokes $V$ parameter), which detected a surface magnetic field in $\Theta^1$~Ori~C, a chemically non-peculiar hot star~\cite{donati2002}.  
Since then, large-scale spectropolarimetric surveys have been conducted to study the magnetic properties of stars with O, B, and A spectral types~
{\cite{petit2013,morel2015,fossati2015,fossati2015b,fossati2016,wade2016,neiner2017,grunhut2017,sikora2019,shultz2019b}}. 
This led to valuable statistical information on magnetism in high-mass stars. The surveys concluded that magnetism is unambiguously detected in about 10 percent of the sample stars, regardless of spectral type. The main common ingredient seems to be the similarity in stellar structure (convective core and a radiative envelope), but dependence on stellar parameters (mass, rotation) could not be found~{\cite{neiner2015,shultz2019b}}.
It has also become clear that chemical peculiarities may not be evidenced in some magnetic stars. The lack of such peculiarities is generally attributed to stellar winds, which can prevent vertical chemical stratification in the upper stellar layers. Indeed, many magnetic O and B-type stars do not exhibit the same kind of peculiarities as Ap and Bp stars, yet their magnetic characteristics are broadly consistent.

The empirical characterisation of these magnetic fields suggests that the typical structure is predominantly an oblique dipole (misaligned magnetic and rotation axes) in the majority of the cases, although extensive spectroscopic and spectropolarimetric monitoring may help reveal departures from the first-order pure dipole fits, e.g.,~\cite{daviduraz2021}. 
%
An important clue about the possible origin of magnetism in high-mass stars relates to recent measurements of obliquity angles by~\cite{shultz2018}. The findings indicate a random distribution in the tilt of the magnetic axis compared to the rotation axis of the star. This also indicates that magnetic precession may have a longer timescale than the main sequence evolution of the star~\cite{lander2017,lander2018}. 
In addition, a small fraction of known magnetic hot stars cannot be reconciled with dominantly dipolar geometries: their peak magnetic field strengths are in higher-order spherical harmonic components, suggesting a complex magnetic geometry, e.g.,~\cite{donati2006,kochukhov2016}. Interestingly, these stars are generally consistent with a young stellar age.

The inferred magnetic field strengths range from a few hundred G to up to tens of kG. The strongest magnetic field reported ($\approx$20~kG inferred dipolar strength) for a seemingly chemically normal O-type star is of NGC 1624-2~\cite{wade2012,daviduraz2021}. 
The detection of ``ultra-weak'' magnetic fields in Vega~{\cite{lignieres2009,petitp2022}}, an intermediate-mass A-type star, has been surprising. A few other A-type stars have been discovered to host ultra-weak fields, although they possess chemical peculiarities typically associated with the Am class~\cite{blazere2016, blazere2020}. It is unknown if such weak magnetism may or may not be a widespread phenomenon since the origin of such fields is unclear. However, the observed strong fields are now commonly explained as fossil remnants~{\cite{moss2003,braithwaite2004,mestel2010,ferrario2015}}. 
The bi-modality of observed magnetic field strengths is separated by a ``magnetic desert'', i.e., a scarcity of inferred dipolar magnetic field strengths in the range from a few G to about three hundred G~\cite{auriere2007,lignieres2014}. It seems that this cutoff separates magnetic fields that may either be produced by different mechanisms or may reflect on the distribution of initial magnetic field strengths~\cite{petit2019}.
The results from spectropolarimetric surveys raise important questions that are major drivers for current investigations: Why do only a fraction of hot stars show evidence of strong magnetic fields? How do these fields evolve over time? How does the magnetic geometry change over time? 
To gain further insights from observations, it is essential to increase the sample size of known magnetic high-mass stars; therefore, high-yield surveys are much desired. A successful strategy has been the identification of characteristic light curve variations in photometric data and subsequent spectropolarimetric follow-up observations, e.g.,~\cite{bram2018,pfeffer2022,shen2023,mobster1}.
Recent advancements include further diagnostics from infrared observations \mbox{, e.g.,~\cite{eikenberry2014,oksala2015,wis2015}}, radio observations, e.g.,~\cite{leto2021,das2022,erba2022,owocki2022,mobster6}, and planned UV spectropolarimetric missions, e.g.,~\cite{morin2019,shultz2022,folsom2022}. The multiwavelength characterisation is invaluable to fully explore the physical behaviour of observed magnetic fields.
One incredibly important piece of missing information, the metallicity dependence of observed magnetism~\cite{bagnulo2017,bagnulo2020}, is beyond reach with current instrumentation. High-resolution optical spectropolarimetric units, suitable for stellar magnetometry of hot stars, are only mounted on 4 m class telescopes. To detect magnetic high-mass stars in the Magellanic Clouds and beyond, it is essential to include high-resolution spectropolarimetric units on 10 m class and larger aperture telescopes.

\subsection{Neutron Stars and Magnetars}\label{sec:magnetar}

Although the detailed discussion of compact objects is beyond the scope of this review, we should outline that there is ongoing debate as to how the magnetic field of these objects is acquired. In particular, a subclass of neutron stars, magnetars, shows evidence of surface magnetic fields of the order of $10^{13}$--$10^{15}$~G, whereas the magnetic field strengths of typical neutron stars range between $10^{11}$ and $10^{13}$~G~\cite{kaspi2017}.
One scenario suggests that the magnetic flux of main sequence stars is conserved and serves directly as the magnetic field of the remaining compact object~\cite{ferrario2008,ferrario2015,zhou2019}. 
A major challenge for this scenario is that magnetic flux should be essentially entirely preserved during stellar evolution and core collapse. It is unclear if these assumptions would hold.
Another issue is that the fraction of strongly magnetised main sequence stars may not match the fraction of strongly magnetised neutron stars~\cite{makarenko2021}. 

Other origin scenarios involve in situ amplification from a seed magnetic field at around the time of core collapse via a dynamo mechanism~{\cite{thompson1993}}. This could be realised via a convective dynamo, e.g.,~\cite{masada2022}, a magneto-rotational instability-driven mechanism, e.g.,~\cite{reboul2021,aloy2021,ober2022}, or even a combination of these~\cite{reboul2022}. New approaches have also considered a Tayler--Spruit mechanism to further amplify a 10$^{12}$~G magnetic field in a few seconds to 10$^{16}$~{G}~\cite{barrere2022}. 

In order to understand the origin of strong magnetic fields in magnetars, it remains an essential task to map out the magnetic characteristics of their progenitors and constrain how much magnetic flux may be available at the time of core collapse. This could also have implications for detections of fast-radio bursts that have been linked to magnetars, e.g.,~\cite{boc2020,chime2020,petroff2022}.

\section{Theoretical Background}

In this section, we overview some key concepts that form the foundation of new modelling approaches. These concern the kinds, stability, and effects of magnetic fields. There are several excellent textbooks and reviews on the topic---for example,~\cite{mestel1999,kulsrud2005,brandenburg2005,goedbloed2010,walder2012,braithwaite2017,augustson2020}.

\subsection{Fossil vs. Dynamo Fields}

Stars could inherit some magnetic flux from the molecular cloud. The parent molecular cloud has its own magnetic field. As the cloud collapses, the magnetic field lines are compressed, and the field strength increases. This amplification is a consequence of the ``frozen-in'' flux condition (see below).
In fact, it could be shown that some magnetic flux must be dissipated (via non-ideal MHD processes) during this phase to avoid a ``flux catastrophe'' problem of ending up with too much magnetic flux in stars~\cite{commercon2011,braithwaite2012}. 
A remnant magnetic field from star formation can relax into an equilibrium state and form a \emph{large-scale} structure. Demonstrating the stability of such fields has been a pioneering work by Braithwaite and collaborators~\cite{braithwaite2004,braithwaite2006,braithwaite2008,braithwaite2009} and has also been recently revisited~\cite{becerra2022,becerra2022b}. 

Magnetic fields induced by a dynamo cycle are fundamentally different from the above picture. A dynamo cycle converts kinetic energy to magnetic energy and thus operates on a completely different time scale. Dynamo theory has been extensively developed; for a comprehensive review, see, e.g.,~\cite{brandenburg2005}.
Several dynamo mechanisms describe a local phenomenon. Therefore, dynamo-produced fields are often characterised by length scales that are relevant to the powering mechanism (e.g., convection). The perturbed magnetic fields can be considered as \emph{small-scale}, for example, in comparison to the size of the star. 
An electromotive force is necessary to reconstruct a large-scale field from small-scale perturbations. This is often described via the \emph{mean-field theory}~\cite{mestel1999}. There are nonetheless still debates about relevant length scales and as such whether some dynamos would be small- or large-scale. 

%
%
\subsection{Time Evolution of Magnetic Fields}

A general MHD equation of the time evolution of magnetic fields can be cast in the form of:
\begin{linenomath}
\begin{equation}\label{eq:dbdt}
    \frac{\partial \mathbf{B}}{\partial t} = \grad \times (\mathbf{u} \times \mathbf{B} + \eta \grad \times \mathbf{B}) \,, 
\end{equation}
\end{linenomath}
where $\mathbf{B}$ is the magnetic field vector, $t$ is the time, $\grad$ is the spatial derivative in three dimensions, $\mathbf{u}$ is the fluid velocity, and $\eta$ is the magnetic diffusivity.

%
\subsubsection{Fossil Fields}
In the case of relaxed, equilibrium, fossil magnetic fields, the above equation simplifies~to: 
\begin{linenomath}
\begin{equation}\label{eq:dbdt_fossil}
    \frac{\partial \mathbf{B}}{\partial t} = \eta \grad^2 \mathbf{B} \,, 
\end{equation}
\end{linenomath}
assuming that the magnetic diffusivity has no spatial dependence. There is no active generation or induction term. 
Assuming that the most relevant physical process leading to the dissipation of fossil fields is Ohmic diffusion, the evolution of fossil fields is expected on a very slow (Ohmic) time scale.
It can be shown that, for the Sun, a general Ohmic dissipation time scale ($\tau_{\rm Ohm} \approx l^2 / \eta$, with $l$ being a characteristics length scale) is longer than the nuclear-burning time scale~\cite{cowling1945}. 
The accurate calculation of Ohmic dissipation is difficult in the context of high-mass stars. The magnetic diffusivity varies as a function of radial distance inside the star and depends on thermodynamic quantities. For example, the temperature profile drops several orders of magnitude from the core to the surface. Thus, current estimates for the magnetic diffusivity remain quite uncertain~\cite{charbonneau2001}. It is necessary to understand how the fossil field is anchored inside the star to obtain better quantitative estimates of dissipation processes. 
In high-mass stars, the nuclear-burning time scale is much shorter than in the Sun ($\sim$ few Myr vs. $\sim$10 Gyr).
If the Ohmic dissipation time scale for the Sun and high-mass stars were roughly similar, any fossil field in high-mass stars should be long-lasting.
Some additional mechanisms or phenomena still cannot be ruled out as potential sources of magnetic field dissipation.
An interesting possibility may be tidal dissipation by a companion star~\cite{vidal2019}. However, rapid dissipation requires a close companion on a short-period (<20 d) orbit. Although multiplicity is common amongst high-mass stars, the number of short-period systems is a more moderate fraction. Thus, it is not immediately evident what fraction of stars are affected by this process.
In general, the dissipation of fossil fields is expected to be slower than the star's evolution. The principle of the frozen-in flux condition (Alfv\'en's theorem~\cite{alfven1942}) can be applied to deduce the surface magnetic field strength. The magnetic flux over a closed surface area $\mathbf{S}$ is defined as:
\begin{linenomath}
\begin{equation}\label{eq:flux_conservation}
  \Phi = \oint_S \mathbf{B} \mathrm{d} \mathbf{S} \approx 4 \pi R^2 B \,, 
\end{equation}
\end{linenomath}
where the unsigned magnetic flux is considered for the surface of a sphere with radius $R$. Since the time derivative of the magnetic flux is zero, the above expression must be constant, requiring a relation between the magnetic field strength and stellar radius.
If we define the surface strength of the magnetic field in a time-dependent manner, we can write: 
\begin{linenomath}
\begin{equation}\label{eq:flux_conservation2}
  B_{\rm surf} (t) = B_{\rm surf} (t_0) \frac{R_{\rm surf}^2 (t_0)}{R_{\rm surf}^2 (t )}
\end{equation}
\end{linenomath}
with $R_{\rm surf}$ being the stellar radius and $t_0$ an initial time (typically assumed to be the Zero Age Main Sequence). Since the radius of the star changes during its evolution, we may anticipate a change in the surface field strength in response. Some observational studies are consistent with this picture~\cite{neiner2017, martin2018,sikora2019}, while others infer a more rapid decline of the magnetic field strength by a magnetic flux decay~\cite{fossati2016,shultz2019b}. 
These approaches often rely on stellar ages inferred from non-magnetic evolutionary models. This source of uncertainty can now be improved upon with models incorporating fossil field effects~\cite{keszthelyi2020,keszthelyi2022}.

%

%
\subsubsection{Dynamo Fields}

Dynamo fields have been proposed in several forms~\cite{kulsrud2005}. 
For example, the induction equation for a convection and rotation-driven, so-called $\alpha$-$\Omega$ dynamo can be written as: 
\begin{linenomath}
\begin{equation}\label{eq:dbdt_conv}
    \frac{\partial \mathbf{B}}{\partial t} = \grad \times (\mathbf{u} \times \mathbf{B} + \alpha \mathbf{B} ) \,,
\end{equation}
\end{linenomath}
where, for simplicity, we assumed no dissipative terms. Such dynamos are mostly constrained by the Rossby number $Ro = 2 \pi / \Omega \tau_{\rm conv}$, with angular velocity $\Omega$ and the convective turnover time scale defined as the ratio of a characteristic length scale and convective fluid velocity $\tau_{\rm conv} = l / v_{\rm conv}$, e.g., recently
~\cite{landin2023}. This implies that such dynamos will show a dependence on the rotation rate of the star.
In Equation~\eqref{eq:dbdt_conv}, the $\alpha$ effect is responsible for turbulent convection, which displaces a toroidal magnetic field and converts it into a poloidal structure. 
The $\Omega$ term (encompassed in the fluid velocity vector $\mathbf{u}$) accounts for shearing the magnetic field by differential rotation, which regenerates a toroidal component from a poloidal one.
Variations of this kind of dynamo include so-called $\alpha^2-\Omega$ and $\alpha^2$ dynamos~\cite{kulsrud2005}. 
The evolution of dynamo fields takes place on a short, \emph{Alfv\'en} time scale, which is characterised by 
$\tau_{A} \approx l / v_{A}$
with $v_{A} = B_{\rm rms} / \sqrt{4 \pi \rho}$ the Alfv\'en speed. Typical time scales in stars are of the order of a few days to a few decades.

%
\subsection{Stability Criteria for Fossil Fields}\label{sec:fossileq}

\begin{figure}[H]
\includegraphics[width=\textwidth/2]{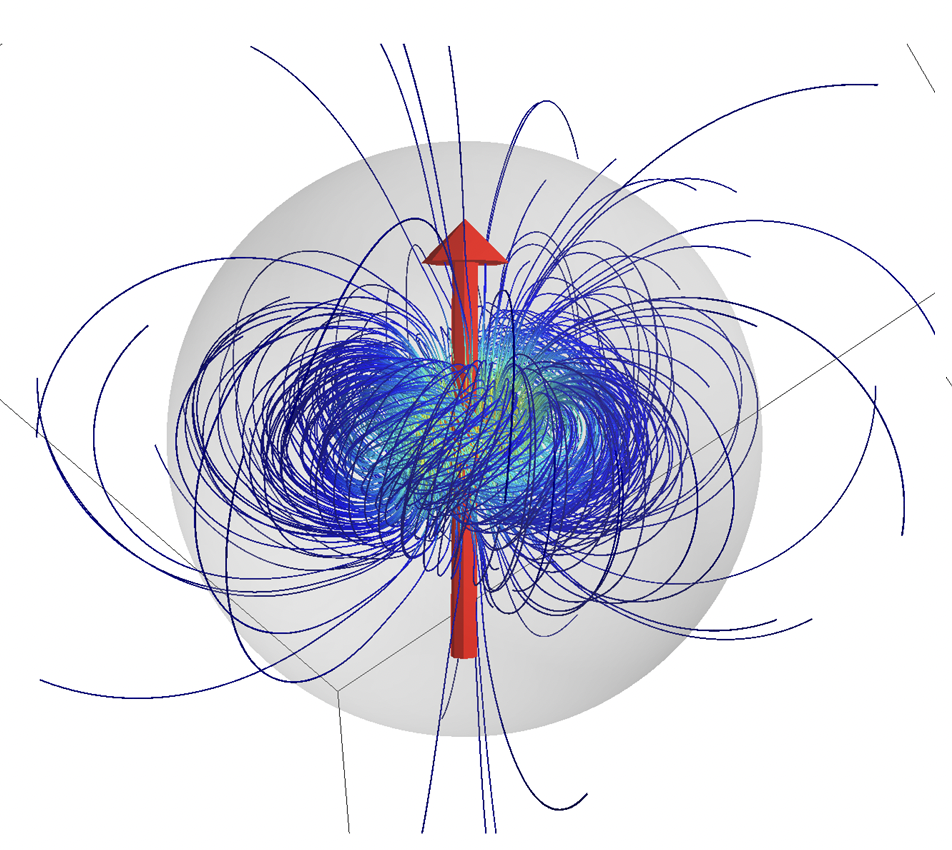}\includegraphics[width=\textwidth/2]{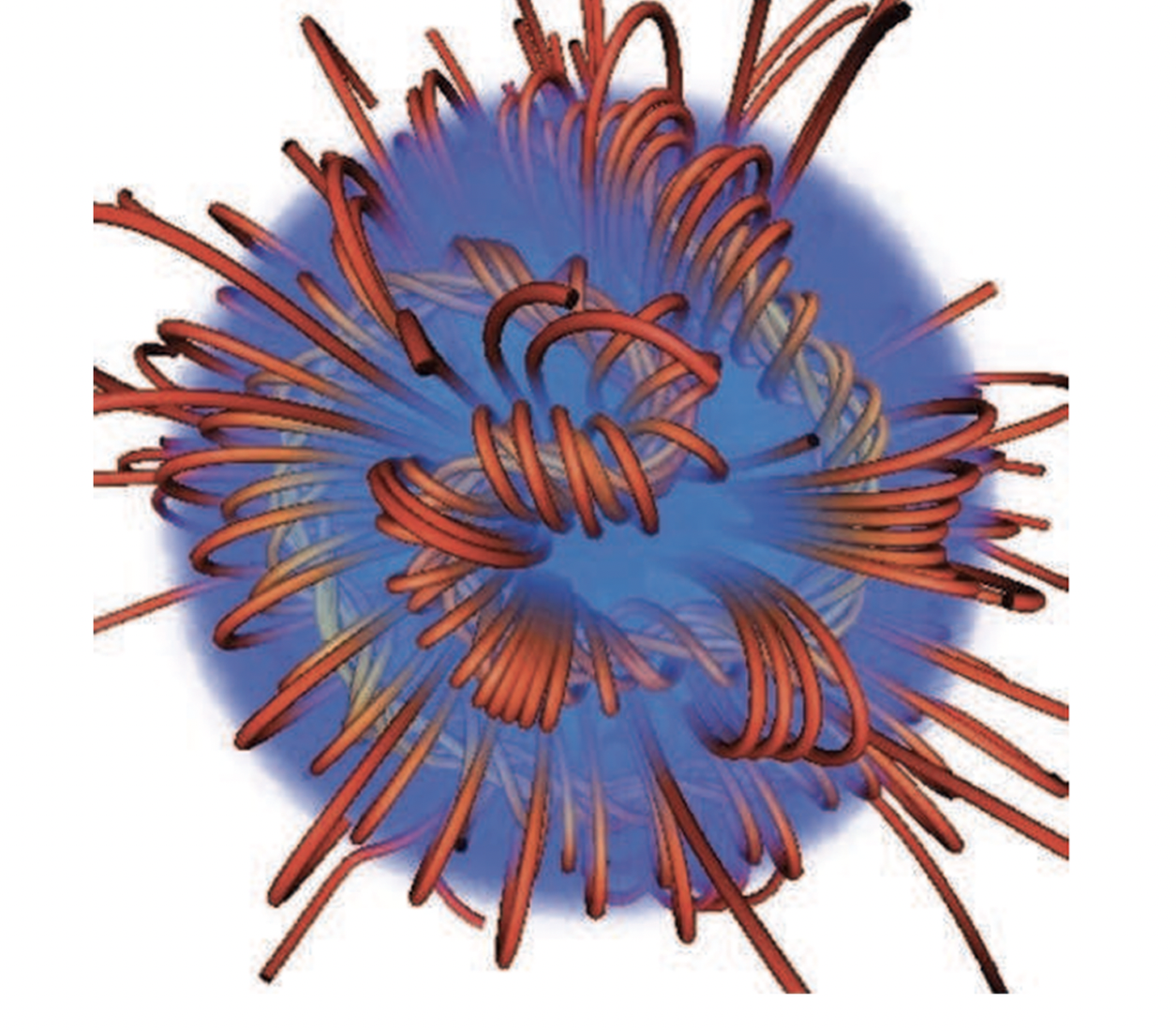}
\caption{Examples of stable fossil field configurations. \textbf{Left}: Axisymmetric dipole field \protect{\cite{becerra2022}}. The fossil field is in stable equilibrium after more than 600 Alfv\'en times. Coloured lines represent magnetic field lines. The grey colour corresponds to the stellar photosphere, and the magnetic axis is shown with the red arrow. \textbf{Right}: Non-axisymmetric, complex magnetic structure in a stable configuration after several tens of Alfv\'en times \protect{\cite{braithwaite2008}}. The blue background represents the star as a spherical gas. Red lines indicate magnetic field lines.}\label{fig:fossil}
\end{figure}

While equilibrium states were traditionally found in several forms, the stability of fossil fields has been more challenging. 
Rigorous studies of the stability of fossil fields have been conducted in numerical simulations~\cite{braithwaite2006,braithwaite2008,braithwaite2009,becerra2022,becerra2022b}. Several stable configurations exist. Figure~\ref{fig:fossil} shows two of them: an axisymmetric dipole and a non-axisymmetric, more complex magnetic structure~\cite{becerra2022,braithwaite2008}. Some stable configurations also exist without any observable surface manifestation.

%
%
These studies confirmed early analytical constraints that both purely toroidal~\cite{tayler73,akgun2013} and purely poloidal~{\cite{flowers1977, markey1973,wright1973}} magnetic field configurations are unstable. Following the analytical model of Prendergast~\cite{prendergast1956}, a stable configuration in numerical models was found by requiring a twisted torus shape of mixed toroidal and poloidal modes, in which the toroidal field is confined by poloidal lines~\cite{braithwaite2004}. In barotropic cases, the magnetic field can be calculated from a Grad--Shafranov type equation~\cite{duez2010,duez2010c}. This concerns the stellar interiors and has also been found to be an equilibrium configuration in the presence of rotation~\cite{yoshida2006,yoshida2006b}. However, the stability of Prendergast-type fields over the course of stellar evolution has been questioned recently~\cite{kaufman2022}.

%
%
%
 
\begin{figure}[t!]
\begin{center}
\includegraphics[width=10cm]{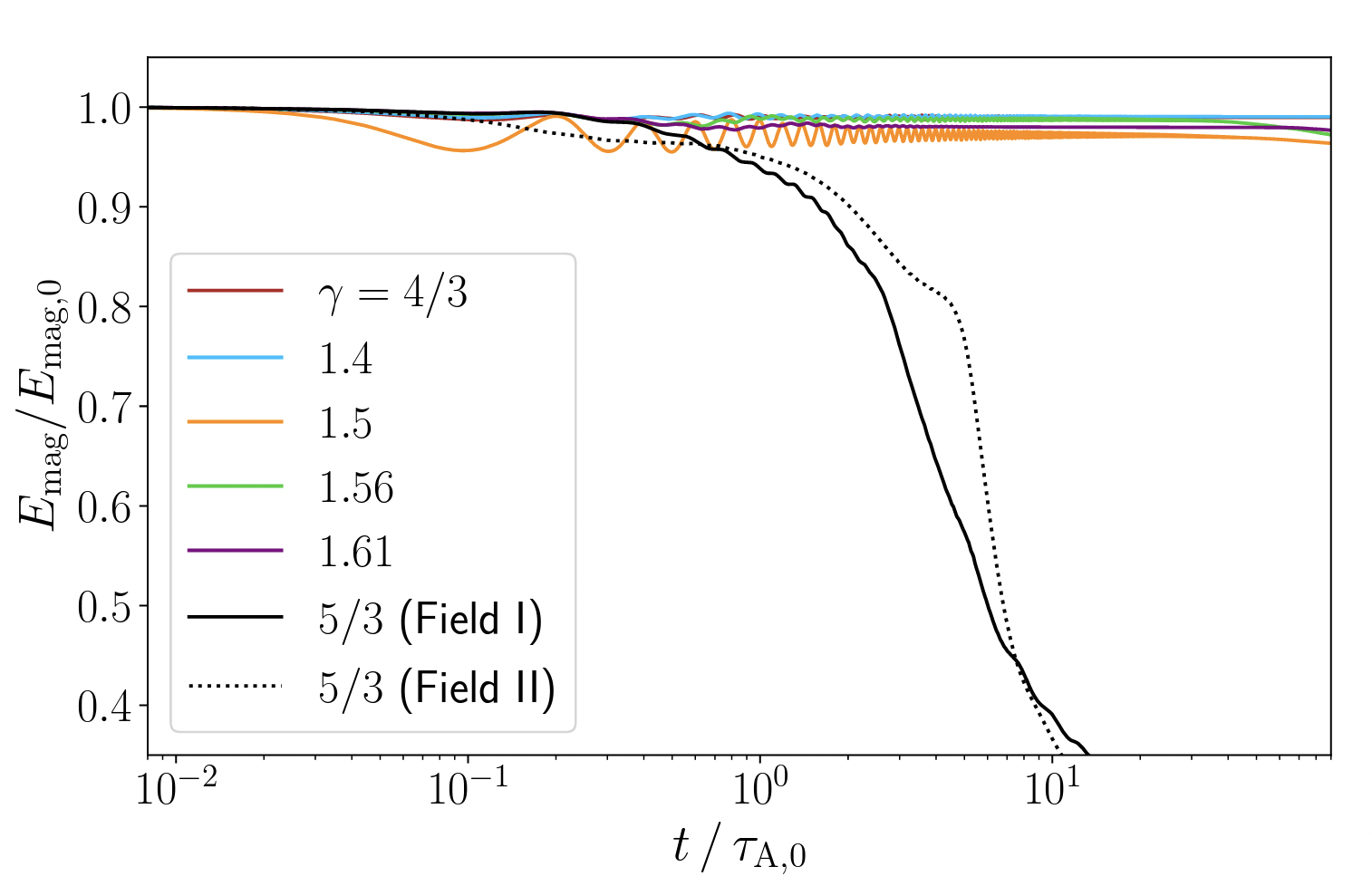}
\caption{Magnetic energy as a fraction of its initial value over Alfv\'en time scales. The polytropic index $\gamma = \mathrm{d} \ln P / \mathrm{d} \ln \rho$ is varied. In a barotropic case, it equals the adiabatic exponent, $\gamma = \Gamma = \ 5/3$. The magnetic energy declines without reaching a stable configuration in this case, from \protect{\cite{becerra2022b}}.}\label{fig:baro}
\end{center}
\end{figure}

%
%
Some controversy also concerns the use of a barotropic equation of state. Using analytical works and numerical simulations, it has been argued that non-barotropic conditions may be important for stable stratification and thus for the stability of the magnetic field; see Figure~\ref{fig:baro}~\cite{reisenegger2009,lander2012,armaza2015,mitchell2015,becerra2022b}. Barotropic refers to regions where the density depends only on the pressure ($\rho = \rho (P)$, or in other words, isobars coincide with iso-density regions) but not on other thermodynamic quantities. 

For a realistic radiative stellar envelope, stable stratification, an outward decreasing density and an increasing specific entropy profile are necessary. According to~\cite{reisenegger2009}, this does require non-barotropic conditions, i.e., knowledge of other thermodynamic quantities, for example, the temperature or entropy in order to remain stable against convection. 

The dipole mode may form a magnetosphere around the star; however, any toroidal field diffuses above the stellar surface such that the field can eventually become force free~{\cite{braithwaite2017,stift2022}}, meaning that the Lorentz force is zero, $(\nabla \times \mathbf{B}) \times \mathbf{B} = 0$. While the magnetic field above the stellar surface can possess an \emph{azimuthal component} (for example, due to stellar rotation shearing a pure dipole field), enclosed toroidal fields above the stellar surface are unstable and rapidly diffusive into the poloidal component.

\subsubsection{Momentum Equation}

Let us write the force balance, the MHD Euler equation of motion, as: 
\begin{linenomath}
\begin{equation}\label{eq:euler}
   0 = - \grad P + \rho \mathbf{g} +  \frac{1}{4 \pi} (\grad \times \mathbf{B}) \times \mathbf{B} + 2 \mathbf{\Omega} \times \mathbf{u}
\end{equation}
\end{linenomath}
where $P$ is the pressure, $\mathbf{g}$ is the gravitational acceleration, $\mathbf{B}$ is the magnetic field vector, $\mathbf{\Omega}$ is the angular velocity, and $\mathbf{u}$ is the fluid velocity. Inside stars, the magnetic field is non-force free, and there is a non-zero Lorentz force.

A fossil magnetic field must be stable against various instabilities; otherwise, it would decay and become unobservable. Here, we describe two interactions that are key to the stability of fossil magnetic fields. In both cases, the stability of the fossil field relies on a critical value. It is thus required for an initial magnetic field to have a critical strength in order to ``survive'' these interactions and relax into a stable equilibrium configuration.

%
%
\subsubsection{Convective Expulsion and Inhibition of Convection}

A stability criterion for magneto-convection, the interplay between convection and magnetism, was first given by~\cite{gough1966}. A first-order estimate can be posited by considering that convection in stellar layers can be suppressed by a critical magnetic field when the magnetic pressure reaches the thermal pressure:
\begin{linenomath}
\begin{equation}\label{eq:equi}
   \frac{1}{2} \rho v = \frac{B^2}{8 \pi} \,, 
\end{equation}
\end{linenomath}
that is, when the magnetic field has equipartition strength, e.g.,~\cite{auriere2007,sundqvist2013}.
Recently, a more sophisticated model was considered, including radiation pressure~\cite{macdonald2019}. The revised stability condition is formulated as:
\begin{equation}\label{eq:magconv}
    Q (\nabla - \nabla_{\rm ad}) < \frac{v_{A}^2}{v_{A}^2 + c_{s}^2} \left(1+ \frac{\mathrm{d}\ln\Gamma_1}{\mathrm{d}\ln P}\right) \,, 
\end{equation}
\noindent where $Q$ is the thermal expansion coefficient, $\nabla$ and $\nabla_{\rm ad}$ are the structural and adiabatic gradients, $v_{A}$ and $c_{s}$ are the Alfv\'en and sound speeds, $\Gamma_1$ is the first adiabatic exponent, and $P$ is the local pressure. 
A minimum Alfv\'en speed $v_{A} = B / \sqrt{4 \pi \rho}$, or equivalently a critical magnetic field strength, is needed for the above relation to be satisfied.
If $B > B_{\rm crit}$, convection can be suppressed or even inhibited, e.g.,~\cite{lydon1995}. This is commonly considered in the context of dynamo fields in convective regions of low-mass stars {(e.g.,~\cite{macdonald2017})}.
On the other hand, if $B < B_{\rm crit}$, convection will expel the magnetic field outside of the convectively unstable region. This is commonly known as convective expulsion, e.g.,~\cite{weiss1966,spruit1979,schussler2001}. 
%

%
%
\subsubsection{Rotational Expulsion and Inhibition of Differential Rotation}\label{sec:rotexp}

Similar to the above constraints, differential rotation could also lead to the local removal of an initial magnetic field if the difference in rotation frequency between adjacent layers is much larger than the Alfv\'en frequency~\cite{spruit1999,braithwaite2017}.  
In this case, the magnetic field may be wound-up such that a high magnetic diffusion effectively dissipates it, leaving no magnetic flux in differentially-rotating regions. 

Once a fossil field is formed, the rotation rate of the star will slow due to magnetic braking. Importantly, strong differential rotation cannot develop in those regions where the field spreads~\cite{ferraro1937,mestel1999}. Instead, the radial rotation profile becomes uniform~\cite{woltjer1959}.
Therefore, models with fossil fields are unlikely to be compatible with strong differential rotation, unless they exist in different regions of a star.

%
%
\subsection{Angular Momentum Transport and Loss}\label{sec:magbr}

Magnetic field lines are very efficient at transporting angular momentum.
We can define a magnetic stress tensor (Maxwell stress tensor) $\mathbf{T}$ as a quantity whose negative divergence yields the Lorentz force: 
\begin{linenomath}
\begin{equation}\label{eq:Trphi}
   \mathbf{F_{\rm Lorentz}} = - \grad \cdot \mathbf{T} \, .
\end{equation}
\end{linenomath}
The Maxwell stress tensor leads to a momentum flux (force per unit area), also known as Poynting flux. The $r, \phi$ component of the stress tensor is given by:
\begin{linenomath}
\begin{equation}\label{eq:Trphi2}
   T_{r, \phi} = \frac{B_r B_\phi}{4 \pi} \,,
\end{equation}
\end{linenomath}
with $B_r$ and $B_\phi$ the radial and azimuthal component of the magnetic field vector, respectively. Integrating this expression over a solid angle yields the azimuthal magnetic force. The scalar value of the magnetic torque is then: 
\begin{linenomath}
\begin{equation}\label{eq:torque}
  N = B_r B_\phi r^2 r \sin\theta \, 
\end{equation}
\end{linenomath}
with $r$ the radius and $\theta$ the co-latitude. This is equivalent to the rate of change in angular momentum per unit time.

There are different scaling relations in the literature to express $B_r$ and $B_\phi$, depending on the problem (for example, the saturation amplitude of a dynamo field) or simply the kind of magnetic field considered. Consequently, expressions for the magnetic viscosity (or resulting magnetic diffusion) can vary, e.g.,~\cite{spruit2002,rudiger2015,fuller2019,schneider2019}. 
In turn, an exact magnetic angular momentum transport equation, which is usually assumed to be a diffusive process, is still somewhat uncertain~\cite{aerts2019}. 
It has been argued that the magnetic Maxwell stress leads to an advective term in the angular momentum transport equation, which could aim to drive the system towards uniform specific angular momentum instead of uniform angular velocity~\cite{potter2012b}.
The diffusivity, whether produced directly via Maxwell stresses or indirectly via magnetic turbulence, must be high. Indeed, it is expected that poloidal field lines could ``freeze'' rotation and enforce solid-body rotation (Section~\ref{sec:rotexp}).
In addition, magnetic fields are also expected to affect wave propagation in stellar interiors, e.g.,~\cite{mathis2012,fuller15,lecoanet2022}.

Considering that there is an outward mass flux from stars,~Weber and Davis~\cite{weber1967} showed that the net rate of change in angular momentum per unit time caused by the solar wind can be cast in the form of: 
\begin{linenomath}
\begin{equation}\label{eq:torque2}
  \frac{\mathrm{d} J }{\mathrm{d} t} = \frac{2}{3} \dot{M} \Omega_\star R_A^2, 
\end{equation}
\end{linenomath}
where $\dot{M}$ is the mass-loss rate, $\Omega_\star$ is the surface angular velocity, $R_A$ is the Alfv\'en radius, and the factor 2/3 results from integrating over the sphere. 
The occurring mass loss implies angular momentum loss, thus the above formula is commonly adopted to describe the stellar spin down via magnetic braking~\cite{mestel1968,parker1963}.
In the Weber and Davis model, $R_A$ is defined as a location where the Alfv\'enic Mach number is unity. In pioneering MHD simulations of massive stars, it was shown that this analytical scaling relation remains a good approximation for the net rate of angular momentum loss, albeit changing the expression for the Alfv\'en radius from the split monopole model to a more appropriate dipole (or more general, higher-order harmonic) scaling~\cite{ud2008,ud2009}. Subsequent improvements concern including the tilt angle of the magnetic field, which only modestly impacts the above scaling relation~\cite{ud2023}.

\section{Modelling Approaches}

\subsection{Dynamos in the Stellar Core}\label{sec:coredynamo}

For the example of solar-type magnetism, it is nowadays commonly expected that the convective cores of massive stars are good candidates to induce a similar $\alpha-\Omega$ dynamo cycle and produce magnetic fields. Such convection and rotation-driven dynamos were first studied via semi-analytical methods and numerical one-dimensional stellar evolution models~\cite{charbonneau2001}.
A central question has been whether magnetic fields produced in the core could reach the stellar surface. However, opposite conclusions were reached regarding the buoyant rise time of magnetic flux tubes, and this question remains unsettled ~\cite{macgregor2003,macdonald2004}. 

Thus far, a few numerical MHD simulations have targeted this domain.
The convective cores of A and B-type star models were studied, revealing strong dynamo activity~\mbox{\cite{brun2005,feat2009,augustson2016}}. 
The first MHD model of a near-core magnetic field of a B-type star, supported by asteroseismic inference, has recently been studied by~\cite{lecoanet2022}. The inferred field strength is unlikely to be reached by a fossil field, supporting a convective dynamo origin.

\subsection{Dynamos in Convective Shells}\label{sec:dynshell}

In recent years, it has become possible to perform multidimensional MHD simulations focusing on silicon/oxygen shell burning in massive stars~\cite{varma2021}. 
This stage is within a few minutes of the star reaching core collapse. The rotational and magnetic properties at this stage can have a large impact on the core dynamics. For example, asphericities may aid the neutrino-driven explosion. The magnetic field might also become the seed of a subsequent dynamo action to create the neutron star's magnetic field.
The inclusion of magnetic fields is a great advancement in such models in addition to previous hydrodynamic simulations~{\cite{meakin2006,mueller2016,yoshida2019,mcneill2022}}.
Such MHD simulations are computationally expensive and still require a comprehensive approach to consider several physical ingredients, for example, stellar rotation. Nonetheless, their impact on compact object formation may have far-reaching consequences.

\subsection{Dynamos in the Stably Stratified Envelope}

The envelopes of high-mass stars on the main sequence are dominated by radiative energy transport. In these stably stratified regions, turbulent convection is absent, hence it cannot support the $\alpha$ mechanism to operate a dynamo.
However, there may exist other ways of supporting a dynamo cycle in radiative stellar regions. A breakthrough in this field took place when Spruit proposed a dynamo that operates via the Tayler instability and differential rotation~\cite{tayler73,spruit2002}. Spruit argues that in general the $\alpha$ effect of convection may be replaced by instabilities in the magnetic field itself~\cite{spruit2002}.
This requires that differential rotation, quantified in the form of $q = \mathrm{d} \ln \Omega / \mathrm{d} \ln r$, must reach a sufficient value. 
In this picture, it is considered that the first instability to set in is the Tayler instability, and it effectively takes the role of turbulent convection as the $\alpha$ mechanism to produce a contemporaneous dynamo cycle~\cite{spruit2002}. \

\subsubsection{Tayler Instability}
The basis of the Tayler instability, also known as Pitts--Tayler instability, is a pinch type deformation of the magnetic field~\cite{tayler73,pitts1985, wright1973}. It was found that a purely toroidal field becomes unstable when reaching a threshold value and starts to produce perturbed fields in the radial and latitudinal directions, that is, diffusing into small-scale poloidal field components.
The Tayler instability is a fundamental piece to understand a possible dynamo cycle, yet it has only recently received some more attention via using analytical methods~\cite{goldstein2019} and 3D simulations~\cite{ji2022}. 

Spruit proposed that a dynamo mechanism could operate by i) differential rotation winding up a poloidal field and generating a toroidal field, and ii) the Tayler instability re-generating the poloidal field~\cite{spruit2002}. It is necessary that the small-scale perturbations are amplified via an electromotive force ($ \mathbf{u} \times \mathbf{B}$) to re-create the poloidal field.
The key point to obtain a stable dynamo cycle is that the amplification must be balanced by some form of dissipation. Amongst other things, this remains a subject of debate. In the work of~\cite{spruit2002}, the dissipation concerns a thermal diffusion of the large-scale toroidal field to saturate the dynamo, and numerical models of~\cite{braithwaite2006} claimed to confirm this cycle.

However, subsequent works have questioned whether the dynamo cycle could operate. In other numerical simulations, it was found that the dynamo cycle does not reach a steady-state since the electromotive force is insufficient to regenerate the large-scale poloidal field~\cite{zahn2007}. 
It was also pointed out that the Tayler instability generates local perturbations; thus, a large-scale dynamo as envisioned, on the length scale of the stellar radius, may not be achieved~\cite{denissenkov2007}. The issue of the length scale on which the Tayler instability operates is still not settled~\cite{aerts2019}. 

Recently, adjustments were made to Spruit's model by supposing that the dissipation concerns the small-scale perturbed fields and not primarily the large-scale toroidal field~\cite{fuller2019}. This way, higher saturation strengths can be reached and consequently the resulting magnetic viscosity is also higher. This allows for spinning down the stellar cores to quite low~rates.

\subsubsection{Magneto-Rotational Instability}

\textls[-15]{Another branch of possible dynamo solutions is related to the magneto-rotational instability (MRI), which was introduced by~\cite{balbus1991} for accretion disks following earlier work of~\cite{chandra1960}. It has also been studied in core-collapse supernova models, \mbox{e.g.,~\cite{masada2012,ober2022,reichert2022,reboul2021,reboul2022}.} Nevertheless, the consideration of this mechanism in main-sequence stellar models has only recently been gaining more momentum~\cite{wheeler2015,jouve2015,jouve2020,griffiths2022}. 
The key principle is that differential rotation must reach a sufficient strength to trigger the instability. A negative angular velocity gradient is required such that the angular velocity increases towards the centre of the star. The MRI is a very rapid exponential magnetic field amplification process that primarily generates a strong toroidal field. It remains a matter of debate whether in realistic stellar conditions the Tayler instability or the magneto-rotational instability could occur first.}

\subsubsection{Mean-Field Dynamos}

Further advances in this field concern dynamo models that differ more substantially from MRI or Tayler-instability driven ones.
In these pictures, the magnetic field is typically derived from the induction equation, and the field generation is attributed to differential rotation via the velocity field of the fluid in a mean-field dynamo~\cite{potter2012b,quentin2018,kissin2018,takahashi2021,petitdemange2022,skoutnev2022}.

\subsubsection{Adaptation in Evolutionary Models}

Despite the various forms and uncertainties of dynamo descriptions, particularly about their saturation strengths and resulting viscosities in stably stratified stellar regions, they are commonly adopted in stellar evolution models of high-mass stars. To a large extent, they result in the favourable outcome that the rotation profile is flattened, e.g.,~\cite{maeder2004}, and the core angular momentum can be impacted to match the spin rates of compact objects. 
The observational verification of these processes is still challenging. The best way to gain more insights may be via asteroseismology of high-mass stars~\cite{bram2018,aerts2019,bowman2020,burssens2020,vanbeeck2020,lecoanet2022,dhouib2022}.

%
%
\subsection{Fossil Field in the Stably Stratified Envelope}

\subsubsection{Magneto-Convection}

Given the vigorous turbulent convection in the cores of main-sequence massive stars, it is likely that fossil fields cannot be relaxed in that region. Early analytical works have practically neglected this constraint and generally assumed that a large-scale magnetic field could settle in a star, permeating through all layers~\cite{flowers1977}. In fact, even more recent simulations assume a self-gravitating non-rotating gas, where the distinct main-sequence structure of a star is not explicitly accounted for~\cite{braithwaite2006,becerra2022}. 
While this constraint was considered in the simulations of~\cite{feat2009}, and in a phenomenological way in the studies of~\mbox{\cite{keszthelyi2022,petermann2015}}, the physical nature of fossil fields near convective cores remains largely unexplored. 

Moreover, hot and high-mass stars on the main sequence may have a very thin layer close to the surface where some small fraction of the total luminosity is transported via convection~\cite{cantiello2009}. This is expected to happen as a result of the iron opacity peak. However, a fossil magnetic field may suppress or completely quench convective turbulence in this region. 
The observational verification of this phenomenon might be possible by studying stellar oscillations. 
OB stars in general show evidence for measurable macroturbulent broadening~\cite{simon2017}. This also includes known magnetic OB stars, where typical macroturbulent values range from 5 to 70 km~s\,$^{-1}$~\cite{grunhut2017,shultz2018}, which on average is lower than in the total sample of OB stars. The only exception is NGC~1624-2 with the strongest measured magnetic field in the sample, where the macroturbulence is between 0 and 3~km~s\,$^{-1}$~\cite{sundqvist2013}. 
The origin of macroturbulence remains debated.
Since fossil fields can inhibit convection by the criterion given in Equation~\eqref{eq:magconv}, it seems challenging to explain that, in these cases, macroturbulence could arise from the sub-surface region.
Instead, it might be more likely that the observed macroturbulence in known magnetic OB stars originates from g-mode pulsations from the stellar core~\cite{aerts2009,bowman2020}, which however could be damped by the magnetic field. The rotational velocities in general are lower in magnetic high-mass stars in comparison to those stars where strong surface fields are not detected. Since macroturbulence may also correlate with stellar rotation, this might explain the lower measured macroturbulent velocities in magnetic high-mass stars.

\subsubsection{Surface Magnetic Braking}

Stars with fossil fields embedded in their radiative envelopes are subject to significant angular momentum loss due to magnetic braking (Section~\ref{sec:magbr}). The spin-down of the stellar surface is clearly evidenced from observations~\cite{oksala2012,town2010,mikulasek2011} and MHD models~\cite{ud2008,ud2009}.

The first massive star evolutionary models taking into account this effect were studied by~\cite{meynet2011} using the Geneva stellar evolution code. They found that models with magnetic braking strongly depend on the assumptions regarding the internal coupling in the star. 
To further investigate this effect, the authors of \cite{keszthelyi2019} also computed Geneva stellar evolution models, including magnetic field evolution (in the form of Equation~\eqref{eq:flux_conservation}, allowing for the surface strength of the magnetic field to change as a function of the stellar radius) and magnetic mass-loss quenching (see below). 
The main conclusion is that magnetic braking can rapidly brake the surface rotation of the star; however, the magnetic properties and the efficiency of rotational braking also change as a function of time. 
Subsequent modelling approaches using the MESA software instrument could investigate a larger parameter space, and hence the dependence on initial mass and initial rotation~\cite{keszthelyi2020}.
Some studies also explored surface magnetic braking in cases where an internal dynamo-driven magnetic field operates~\cite{potter2012b,quentin2018,takahashi2021}. 

Another observable outcome of magnetic braking concerns the mixing of chemical elements. As the star spins down, the time scale of magnetic braking has to be longer than the time scale of chemical mixing to see any surface enrichment of core-produced materials, $^{14}N$ specifically. Quantitatively, this depends on the initial magnetic field strength, its time evolution, and further assumptions about the angular momentum distribution. Correlations have indeed been difficult to identify in observed samples of magnetic stars, e.g.,~\cite{aerts2014,martins2012,martins2015}. While the direct impact of the magnetic field on chemical element transport is not fully clear, stellar evolution models incorporating magnetic braking have shown that the spin-down indirectly impacts the surface chemical composition. In particular, magnetic models can produce slow-rotating nitrogen-enriched stars with high surface gravities, which is not possible with standard evolutionary models otherwise~\cite{keszthelyi2019,keszthelyi2022}. The existence of such stars is well-documented from large-scale spectroscopic studies in the Magellanic Clouds, e.g.,~\cite{lennon2003,ramireza2015,dufton2018,dufton2020}; however, their magnetic characterisation still awaits (see Section~\ref{sec:magobs}).

\subsubsection{Mass-Loss Quenching}\label{sec:quench}

In pioneering MHD models it was shown that the stellar magnetosphere plays a crucial role in the dynamics of the outflow from hot massive stars~\cite{ud2002}. Importantly, channelling of wind material along magnetic field lines leads to shocks and X-ray emission in the magnetic equator~\cite{naze2014}. Subsequent analytical and numerical magnetospheric models also evidenced that the mass trapped in the magnetosphere eventually falls back onto the stellar surface~\cite{owocki2004,owocki2016,townsend2005,ud2008,ud2009,bard2016}, while break-out events may release some plasma~\mbox{\cite{shultz2020,owocki2020,owocki2022}}. 
The net effect is that the total mass outflow is largely reduced by magnetospheric dynamics. The phenomenon is also clearly evidenced in terms of the lower derived stellar mass-loss rates~\cite{driessen2019b}.
Considering the long-term impact of such magnetic mass-loss quenching, stellar evolution models could be studied for the first time accounting for reduced mass-loss rates attributed to an evolving dipolar magnetic field \mbox{\cite{keszthelyi2017a,keszthelyiphd}}. 
The evolutionary impact of magnetic mass-loss quenching is modest for stars with weaker winds, that is, stars initially less massive than about 20~M$_\odot$ at solar metallicity~\cite{keszthelyi2019,deal2021}. However, more massive stars can retain close to all of their initial mass if a strong magnetic field is present~\cite{keszthelyi2020,keszthelyi2022}.
This approach led to subsequent applications in the context of heavy-stellar mass black holes, pair-instability supernovae, and luminous blue variables~\cite{petit2017,georgy2017,groh2020}. 

\begin{figure}[ht!]
\includegraphics[width=\textwidth]{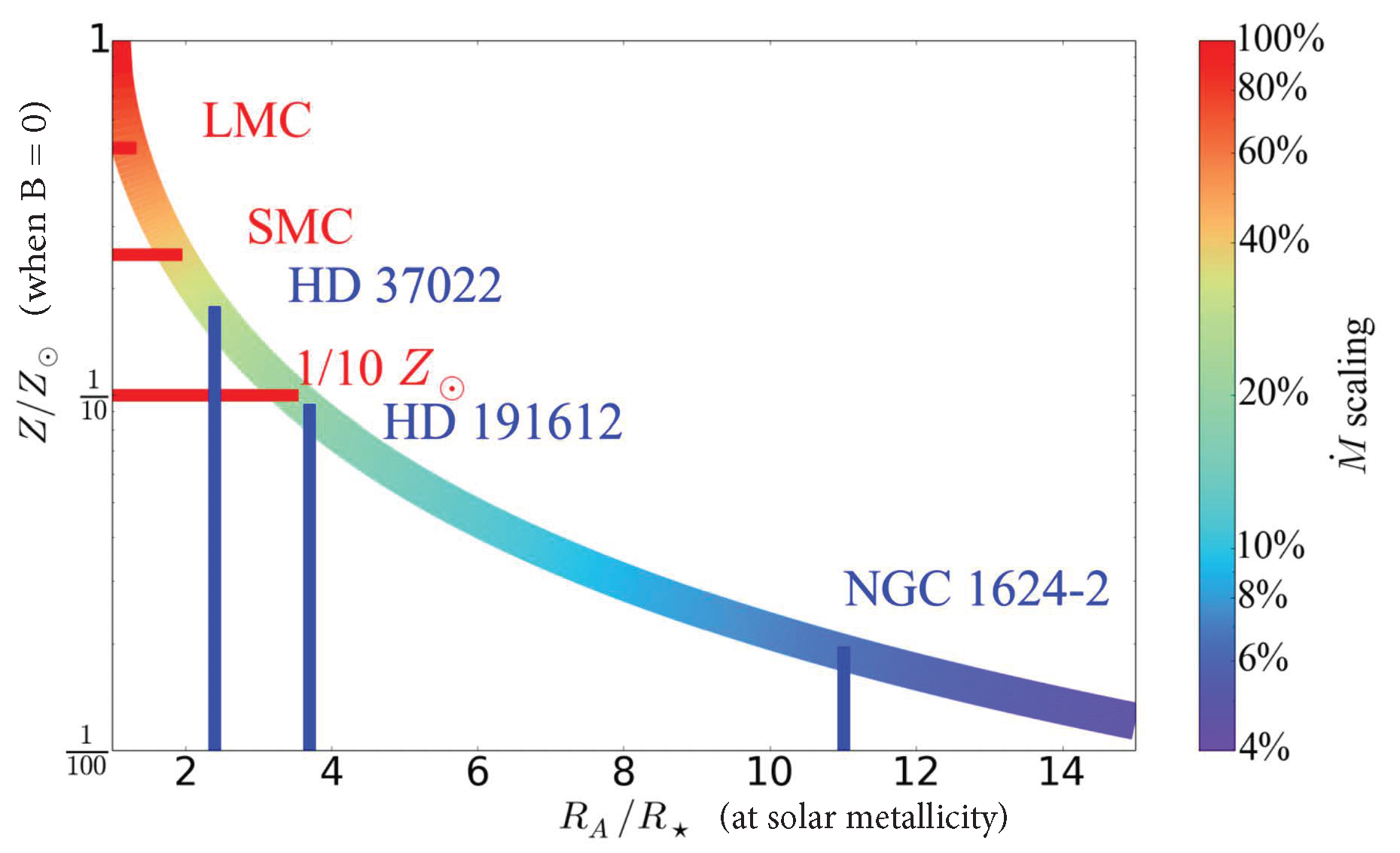}
\caption{Equivalency curve between different physical processes leading to a reduction in stellar mass-loss rates (colour-coded). The horizontal axis shows the Alfv\'en radius (in units of the stellar radius), quantifying the strength of the magnetic field. It is considered at solar metallicity. Some known magnetic massive stars at solar metallicity are included for reference. The vertical axis shows the effect of metallicity for a non-magnetic case. For reference, the Large and Small Magellanic Clouds (LMC and SMC, respectively) are indicated. The first LIGO detection was inferred at 1/10 of solar metallicity. Adopted from Figure 7.2 of \protect{\cite{keszthelyiphd}}.}\label{fig:magmet}
\end{figure}   

%
The key principle is illustrated in Figure~\ref{fig:magmet}. The winds of hot stars are radiatively-driven, primarily due to iron-group elements, e.g.,~\cite{cak1975,pauldrach1986,vink2001,puls2008}.
Mass-loss rates can be impacted by various processes and factors. Considering a reference non-magnetic model at solar metallicity (top left corner of the diagram), its mass-loss rate can be decreased by two independent processes. Lower metallicity leads to a reduction in $\dot{M}$ since there are fewer metals to drive the wind. A typical metallicity scaling is estimated to be $\propto Z^{m}$ with $m$ in the range of 0.70--0.85, e.g.,~\cite{mokiem2007}. This applies not only to OB star winds but also to Wolf--Rayet type winds (which can take away a large amount of mass prior to core collapse) that may be largely impacted by metallicity effects~\cite{sander2020,sander2020b,higgins2021}.
Stronger magnetic fields lead to a reduction in $\dot{M}$ since more mass can be trapped in the magnetosphere and channelled back onto the surface. This can be quantified with a scaling of $\sim 1 - \sqrt{1 - 1/R_c}$ with the closure radius proportional to the Alfv\'en radius $R_c \approx R_\star + 0.7 (R_A - R_\star) $ for a non-rotating dipolar case~\cite{ud2008,ud2009}.
The implication, therefore, is that astrophysical phenomena that were thought to be restricted to low metallicity environments, due to requiring low mass-loss rates, might be possible at high metallicity if the progenitor star is strongly magnetised.

\subsection{Stellar Mergers} 

Stellar mergers are exciting astrophysical phenomena. Although these dynamical events have not been directly observed yet, there are interesting candidates for post-merger objects, e.g.,~\cite{pauldrach2012,hirai2021,vigna2022}. 
The first works suggesting stellar mergers as the origin of observed magnetism in high-mass stars were by~\cite{ferrario2009,wickramasinghe2014}. Subsequently, the authors of \cite{schneider2016,schneider2019,schneider2020} scrutinised whether main-sequence mergers could explain the apparent rejuvenation of stars, developed state-of-the-art MHD simulations, and followed the evolution of the merger product. 
In the pioneering MHD model, a seed magnetic field is greatly amplified via the magneto-rotational instability during the merger of two stars. This is a very promising channel to explain how some stars may become strongly magnetised, and there is support for such a channel for the example of compact objects~\cite{ferrario2015,caiazzo2021}.

The nature of the resulting magnetic field is not immediately clear. The magneto-rotational instability creates a primarily toroidal magnetic field, which is likely close to being perpendicular to the rotation axis, suggesting that there might be some preferential alignment of the resulting magnetic geometries. 
Given the immense computational cost of such simulations, the magnetic field evolution could be followed for 5 Alfv\'en crossing time scales~\cite{schneider2019}. This is much less than the order of few 10--100 Alfv\'en crossing time scales studied for the relaxation of fossil fields~\cite{braithwaite2004,braithwaite2006,becerra2022,becerra2022b}. It is thus not yet fully clear if the merger-induced field amplification actually does relax into a stable fossil configuration. This might sensitively depend on how fast the outer stellar envelope can become stable against convection and differential rotation. 

It was claimed that the models broadly represent the properties of the B-type star $\tau$~Sco~\cite{donati2006,braithwaite2008} as a proof of concept~\cite{schneider2019,schneider2020}. As they noted, some observables were not well matched with the post-merger models, particularly the observed nitrogen abundance and very slow surface rotation of the star. Single-star evolutionary models were studied focusing on the reconciliation of these mismatches~\cite{keszthelyi2021}. They found that order of magnitude changes are required in efficiency parameters to speed up magnetic braking and chemical mixing. This is indeed unlikely to be compatible with a single-star scenario and could perhaps be better explained via a binary channel. 
Nonetheless, it remains difficult to assess if the special case of $\tau$~Sco could represent other early-type stars in general. For example, the magnetic field of this star is much more complex than that of typical OB stars, e.g.,~\cite{kochukhov2016,shultz2018}. 

Observational evidence suggests that main sequence stellar mergers are unlikely to be responsible for generating all fossil fields in high-mass stars. Many known magnetic massive stars are incompatible with such an evolutionary history; in particular, the existence of some young binary systems on stable, nearly circular orbits with one magnetic component, e.g.,~\cite{mobster3} and the doubly magnetic $\epsilon$~Lupi system~\cite{shultz2015} pose a significant challenge.
Some statistical arguments also point toward a discrepancy to explain the origin of fossil fields solely via merger events. The incidence rate of binarity is expected to increase towards higher masses~\cite{mo2017}. Therefore, the occurrence of stellar merger events is also expected to be mass-dependent, with increasing rates towards higher masses~\cite{demink2013}. On the other hand, the incidence rate of observed magnetism is approximately constant across OBA spectral types and inferred masses (see Section~\ref{sec:magobs}).
Moreover, if merger remnants were all strongly magnetised, then certain groups of stars (for example, blue stragglers) should show evidence of high magnetic incidence rates. This is however not observed~\cite{mathys1988,grunhut2017}.

In summary, the merger of two main-sequence stars represents a possible channel to form large-scale fields in high-mass stars. However, it remains disputed whether it could be a general explanation for the origin of fossil fields in all high-mass stars. Ample observational evidence (magnetised pre-main sequence stars, randomly distributed obliquity angles, magnetised components in close binary systems) points toward the direction that fossil fields must remain from earlier stages of the star's history. Thus, the exact origin of fossil fields in high-mass stars is still an unresolved problem.

%
%
%

\section{Overall Picture of Magnetic Field Evolution in a High-Mass Star}

\begin{figure}[H]
\includegraphics[width=13cm]{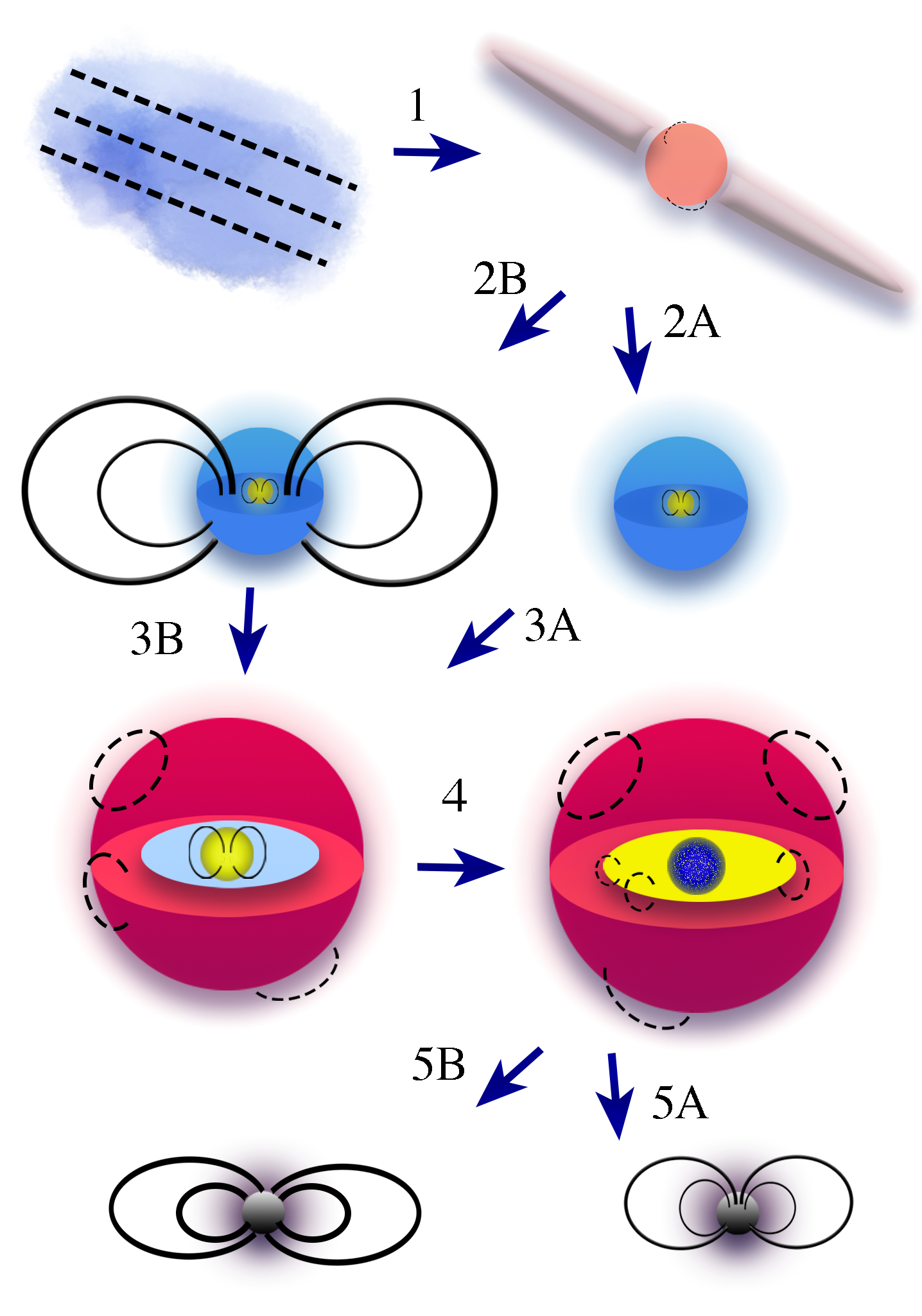}
\caption{Simplified phenomenological picture of magnetic field evolution from early to late stages of stellar evolution for a typical high-mass star. The objects are not to scale. The time evolution proceeds from the molecular cloud phase towards the compact remnants. The black lines indicate magnetic field lines. For stars: Convective burning regions are shown in yellow. Convective non-burning regions are shown in red. Radiative regions are shown in blue. The distinct structure of the pre-supernova Red Supergiant is greatly simplified. See text for the description of the phases, labelled from 1 to 5.
}\label{fig:roadmap}
\end{figure}

Let us now take a simple, qualitative, and phenomenological overview picture of how magnetic fields evolve through stellar evolution, from early to late stages. This concerns various spatial and temporal scales, and many details remain highly uncertain.
Figure~\ref{fig:roadmap} shows a few stages of interest. We can assume that the star is a typical high-mass star of around 20~M$_\odot$, and it evolves through a Red Supergiant phase to core collapse.
%
%
%
\subsection*{Phase 1: Collapse of the Molecular Cloud } 

Molecular clouds are magnetised environments. During their collapse, magnetic fields are wound-up in a disk around the forming star. The proto-star accretes material---and perhaps magnetic flux---from the surrounding magnetised disk.
Dynamo-driven magnetic fields may be initiated in situ in developing stellar convective regions and ``freeze out'' to radiative regions.
Thus far, the magnetic properties of the forming star are very uncertain. 

%
%
%
\subsection*{Phase 2A: Observably ``Non-Magnetic'' OB Star} 

As a main sequence star forms, its structure is relaxed into a convectively unstable core and a stably stratified envelope. The convective core is shown to induce a dynamo action in numerical simulations of A- and B-type star models (Section~\ref{sec:coredynamo}); however, this phenomenon is not yet observationally evidenced. 

The radiative envelope could host a fossil magnetic field without any surface manifestation (Section~\ref{sec:fossileq}). 
If initially differential rotation is present in the system, a dynamo action can also be realised. 
Dynamo-produced magnetic fields are not expected to have any significant surface components either.
Magnetic fields (either fossil or dynamo) in radiative stellar envelopes have not been directly verified via observations. Their inference relies on asteroseismic studies or much more indirect and uncertain effects, for example, the overall spin evolution of the star. These stars, with a lack of direct detection of surface magnetic fields, are often termed ``non-magnetic'', although some form of magnetism presumably exists in their interiors. 
%

%
%
%
\subsection*{Phase 2B: ``Magnetic'' OB Star} 

In a ``magnetic'' massive star, there is evidence for a strong surface component of the magnetic field, as well as co-rotating wind material confined in the magnetosphere (Sections~\ref{sec:magobs} and~\ref{sec:quench}). The magnetic field is most likely of fossil origin. It must be embedded inside the star, satisfying any long-term stability criteria. While the fossil field permeates most or all of the stellar envelope, it is unlikely that it could remain stable in the convective core. In fact, a core dynamo could well take place in such stars, with a complex interaction at the core boundary~\cite{feat2009}. The fossil field enforces rigid rotation in those stellar layers where present; hence, the source of dynamo action in stably stratified regions is~absent.

\subsection*{Phase 3A: Red Supergiant with a Convective Core}

As the OB star finishes core hydrogen burning, it becomes a Red Supergiant (RSG). In the core of the RSG, helium burning will take place. The stability of the core against convection depends on the mass, rotation, and metallicity. A convective helium core might induce a dynamo action for the example of other convective dynamos. Differential rotation in stably stratified regions could also initiate a dynamo cycle. Such internal properties of RSGs remain speculative and poorly constrained. 

%
%

\subsection*{Phase 3B: Red Supergiant with a Convective Core} 
RSGs from ``magnetic'' progenitors could preserve weak (mostly dipolar) fossil fields at the stellar surface~\cite{keszthelyi2019}, if those are able to reconnect along the newly developed, large convective cells. 
The convective surface layers are a likely source of dynamo action~\cite{freytag2002}, which could manifest in weak and complex magnetic fields as observed, for example, in Betelgeuse and some other RSGs, e.g.,~\cite{mathias2018,tessore2017}. 

\subsection*{Phase 4: Advanced Shell Burning Stage} 

As the star undergoes various burning stages, the structural changes sensitively impact the magnetic properties. Fossil fields are essentially unexplored in detail in these stages. The convective silicon--oxygen shell is shown to induce a strong dynamo action in numerical simulations (Section~\ref{sec:dynshell}). Although the changes in the core are rapid, the stellar surface is possibly still characterised by a weak dynamo activity. Models and observations are still scarce to draw firm conclusions.

\subsection*{Phase 5A: Formation of a Typical Neutron Star} 

The magnetic field of a neutron star is either generated at around the time of core collapse or inherited from the precursor. A generating mechanism is not yet clear, but it is possible that neutrino-driven convection or crust formation in the neutron star are key~components. 

\subsection*{Phase 5B: Formation of a Magnetar} 

Similarly to the uncertainties of how typical neutron stars obtain their magnetic fields, the field origin in their subclass of magnetars is also unresolved (Section~\ref{sec:magnetar}).
Arguments based on magnetic flux conservation suggest that OB stars with fossil fields could be the precursors. 
Dynamo models need to invoke a rapid field amplification to reach an order of 10$^{15}$~G strength and saturate at that amplitude. This requires a rapidly-rotating progenitor, which favours a weakly magnetised or rotationally weakly coupled core--envelope configuration. In that case, the progenitor can preserve a larger core angular momentum reservoir. Alternatively, convection or fallback could also power a dynamo mechanism. It is at the forefront of current research to solve these open questions.

\section{Conclusions}

Stellar magnetism has been a much studied area of astrophysics. There is now renewed interest in this field given the understanding that it might drastically impact all stages of stellar evolution and the formation of compact objects. The impact on stellar populations and galactic archaeology is still largely unexplored. 
Major progress has been possible with the advent of extensive photometric, spectropolarimetric, and multiwavelength spectroscopic observations. 
It is essential to continue these efforts and survey extragalactic high-mass stars. Direct magnetic field detection via the Zeeman effect requires consideration of instrument design and equipment of the largest optical telescopes with a high-resolution spectropolarimetric unit.
This could help study the metallicity dependence of magnetism and answer the puzzling origin of observed magnetic fields in high-mass stars.
The detected magnetic fields in main sequence high-mass stars are understood to be of fossil origin. Their impact on stellar evolution has only just begun to be characterised in detailed model calculations. The first large-scale grid of evolutionary models, structure models, and isochrones incorporating fossil field effects of massive stars has recently been published~\cite{keszthelyi2022} and made available at: \url{https://zenodo.org/record/7069766}~\cite{zeno1}. 
Extensive work has focused on dynamo mechanisms in radiative stellar layers. Such dynamo models rely on one of three major pillars: the Tayler instability, the magneto-rotational instability, and the mean-field theory. Thus far, the verification of these models is difficult; however, they may play a crucial role in describing the star's angular momentum evolution from formation to core collapse.
Although dynamos in convective stellar layers are likely to exist, they remain almost completely unexplored in several phases of stellar evolution.
The open questions in this research field motivate further investigations. Magnetic fields can largely impact the physics and evolution of massive stars from their formation until core collapse. Therefore, it remains a contemporaneous topic in stellar astrophysics.

\vspace{6pt} 




\funding{This research received no external funding.}

\dataavailability{Zenodo record of an open library of stellar evolution models accounting for fossil field effects \url{https://zenodo.org/record/7069766} (accessed on 1 March 2023).} 

\acknowledgments{Feedback on the manuscript from Alexandre David-Uraz, the Special Issue Editors, and the Reviewers is greatly appreciated.
}

\conflictsofinterest{The author is a contributor of extensions in \textsc{MESA} that incorporate magnetic field effects (mass-loss quenching and magnetic braking) in evolutionary models.} 






\begin{adjustwidth}{-\extralength}{0cm}

\reftitle{References}




\PublishersNote{}
\end{adjustwidth}
\end{document}